\documentclass[pss,fleqn,embeddedheads]{w-art}
\usepackage{times}
\usepackage{w-thm}

\usepackage[]{graphicx}
\chardef\bslash=`\\ 

\begin{document}

\DOIsuffix{theDOIsuffix}

\Volume{XX} \Issue{1} \Month{01} \Year{2003} 
\pagespan{1}{} 
\Receiveddate{15 November 2003}
\Reviseddate{30 November 2003}
\Accepteddate{2 December 2003}
\Dateposted{3 December 2003}


\title[]{Ferrimagnetism and antiferromagnetism in half-metallic Heusler alloys}


\author[I. Galanakis]{Iosif Galanakis\footnote{Corresponding
     author: e-mail: {\sf galanakis@upatras.gr}, Phone +30\, 2610\, 969925,
Fax +30\, 2610\, 969368}\inst{1}} \address[\inst{1}]{Department of
Materials Science,  University of Patras,  GR-26504 Patra, Greece}

\author[K. \"Ozdo\u{g}an]{Kemal \"Ozdo\u{g}an\inst{2}}
\address[\inst{2}]{Department of Physics, Gebze Institute of
Technology, Gebze, 41400, Kocaeli, Turkey} 

\author[E. \c{S}a\c{s}\i{o}\u{g}lu]{Ersoy \c{S}a\c{s}\i{o}\u{g}lu\inst{3,4}}
\address[\inst{3}]{Institut f\"ur Festk\"orperforschung, Forschungszentrum
J\"ulich, D-52425 J\"ulich, Germany}
\address[\inst{4}]{Fatih University, Physics Department, 34500, B\" uy\" uk\c
cekmece,  \.{I}stanbul, Turkey}

\author[B. Akta\c{s}]{Bekir Akta\c{s}\inst{2}}

\begin{abstract}
Half-metallic Heusler alloys are among the most promising
materials for future applications in spintronic devices. Although
most Heusler alloys are ferromagnets, ferrimagnetic or
antiferromagnetic (also called fully-compensated ferrimagnetic)
alloys would be more desirable for applications due to the lower
external fields. Ferrimagnetism can be either found in perfect
Heusler compounds or achieved through the creation  of defects in
ferromagnetic Heusler alloys.
\end{abstract}

\maketitle

\section{Introduction :}

The family of the ferromagnetic Heusler alloys, e.g. NiMnSb or
Co$_2$MnSi, have been extensively studied during the last years
due to their potential applications in magnetoelectronic devices
\cite{book}. Their main advantage with respect to other
half-metallic systems is their structural similarity with the
binary semiconductors and their high Curie temperatures. First
principles calculations have been extensively employed to study
their electronic and magnetic properties (see Refs.
\cite{Review1,Review2,Review3} and references therein). One of the
most important features of these alloys is the Slater-Pauling
behavior of their total spin magnetic moment which is given simply
as a function of the number of valence electrons in the unit cell
\cite{Gala-Half,Gala-Full}. Authors have studied in the recent
years several aspects of these half-metallic alloys like the
properties of surfaces \cite{Surf1,Surf2,Surf3} and interfaces
with semiconductors \cite{Inter1,Inter2}, the quaternary
\cite{GalanakisQuart,JAP}, the orbital magnetism
\cite{Orbit1,Orbit2}, the effect of doping and disorder
\cite{APL,PRB1}, the exchange constants \cite{ExchConst} and the
magneto-optical properties \cite{Gala-XMCD}.

Half-metallic ferrimagnetic Heusler alloys like Mn$_2$VAl, where
Mn and V atoms have antiparallel spin moments, are of particular
interest since they create smaller external magnetic fields and
thus lead to smaller energy losses \cite{Mn2VZ,Mn2VZ-2}. In the
extreme case like Cr$_2$MnSb (alloys with 24 valence electrons) Cr
and Mn spin moments cancel each other and the compounds are named
as fully-compensated half-metallic ferrimagnets or simply
half-metallic antiferromagnets \cite{Cr2MnZ}. Defects in these
alloys show a very interesting behavior. When we substitute Co
atoms with Cr(Mn) in the Co$_2$Cr(Mn)Al and Co$_2$Cr(Mn)Si
compounds, the impurity atoms couple antiferromagnetically with
the other transition metal atoms lowering the total spin moment
\cite{PSS-RRL,unpublishedSSC}. Co and Fe impurities in Mn$_2$VAl
and Mn$_2$VSi ferrimagnetic alloys have spin moments antiparallel
to Mn and thus the total spin moment reaches closer to the zero
value \cite{Mn2VZ3}. In this manuscript we will complete these
studies presenting results for the Cr(Mn) impurites in
ferromagnetic Co$_2$Mn(Cr)Al and Co$_2$Mn(Cr)Si alloys,  the case
of V and Cr impurities in ferrimagnetic Mn$_2$VAl and Mn$_2$VSi
alloys and finally  the 24-valence half-metallic antiferromagnetic
alloys Cr$_2$FeZ (Z= Si, Ge, Sn) using the full--potential
nonorthogonal local--orbital band structure scheme (FPLO)
\cite{koepernik}.

\begin{table}
\caption{Total and atom-resolved spin magnetic moments in $\mu_B$
for the [X$_{1-x}$X$^*_x$]$_2$YZ compounds. The atom-resolved spin
moments have been scaled to one atom. We do not present the
moments of the Z sp-atom since they are negligible with respect to
the X and Y transition-metal
atoms.}\label{table}\renewcommand{\arraystretch}{1.5}
 \begin{tabular}{lcccc|lcccc} \hline \hline
Compound  & $m^\mathrm{Co}$  &  $m^\mathrm{X^*}$  & $m^\mathrm{Y}$
 & $m^{Total}$ & Compound  & $m^\mathrm{Mn}$  &
$m^\mathrm{X^*}$  & $m^\mathrm{V}$  & $m^{Total}$  \\

[Co$_{0.9}$Cr$_{0.1}$]$_2$MnAl  & 0.72 & -2.57 & 2.79 & 3.43 &
[Mn$_{0.9}$V$_{0.1}$]$_2$VAl  & -1.68 & -1.00 & 1.03  & -2.13 \\

[Co$_{0.9}$Cr$_{0.1}$]$_2$MnSi  & 0.97 & -1.55 & 3.03 & 4.40 &
[Mn$_{0.9}$V$_{0.1}$]$_2$VSi  & -1.17 & -1.10 & 0.92 & -1.33
\\

[Co$_{0.9}$Mn$_{0.1}$]$_2$CrAl  & 0.74 & -1.75 & 1.70& 2.60 &
[Mn$_{0.9}$Cr$_{0.1}$]$_2$VAl  & -1.66 & -1.8 & 1.12  & -2.20
\\

[Co$_{0.9}$Mn$_{0.1}$]$_2$CrSi  & 0.94 & -0.95 & 2.14 & 3.60 &
[Mn$_{0.9}$Cr$_{0.1}$]$_2$VSi  & -1.06 & -1.58 & 0.96  & -1.20
\\

\hline \hline
\end{tabular}
\end{table}

\section{[Co$_{1-x}$Cr$_{x}$]$_2$MnZ  and [Co$_{1-x}$Mn$_{x}$]$_2$CrZ (Z= Al, Si) alloys :}\label{sec1}

In the first part of our study we will concentrate on the case of
defects in the strong ferromagnetic half-metallic Co$_2$CrZ and
Co$_2$MnZ (Z= Al or Si) Heusler alloys. In Ref. \cite{PSS-RRL} we
have studied the case of Cr defects in Co$_2$CrAl and Co$_2$CrSi
alloys; Cr atoms substitute Co atoms at the perfect X sites. We
expanded this study in Ref. \cite{unpublishedSSC} studying the
case of Mn impurities in Co$_2$MnAl and Co$_2$MnSi compounds. In
all cases the impurity atoms had spin moments antiparallel to the
other transition metal atoms lowering the total spin moment while
keeping the half-metallic character of the parent compounds. We
have expanded this study now considering Mn impurities in
Co$_2$CrAl and Co$_2$CrSi and Cr impurities in Co$_2$MnAl and
Co$_2$MnSi and present both the total and atom-resolved DOS's in
Fig. \ref{fig1}. As was the case also in Refs.
\cite{PSS-RRL,unpublishedSSC} all compounds with defects present
the half-metallic behavior and only in the case of
[Co$_{0.9}$Cr$_{0.1}$]$_2$MnSi alloy is the width of the gap
slightly smaller with respect to the perfect Co$_2$MnSi alloy (not
shown here). The total spin moments (see Table \ref{table}) are
considerably smaller than the spin moments of the perfect
compounds : 5 $\mu_B$ for Co$_2$MnSi, 4 $\mu_B$ for Co$_2$MnAl and
Co$_2$CrSi, and 3 $\mu_B$ for Co$_2$CrAl. Thus these kind of
defected alloys could be valuable for realistic applications. As
for the compounds in the previous paragraph we have performed
calculations for several concentrations but results are similar to
the concentration which we present.

\section{[Mn$_{1-x}$V$_{x}$]$_2$VZ  and [Mn$_{1-x}$Cr$_{x}$]$_2$VZ (Z= Al, Si) alloys :}\label{sec3}

\begin{figure}[htb]
\begin{minipage}[t]{.45\textwidth}
\includegraphics[width=\textwidth]{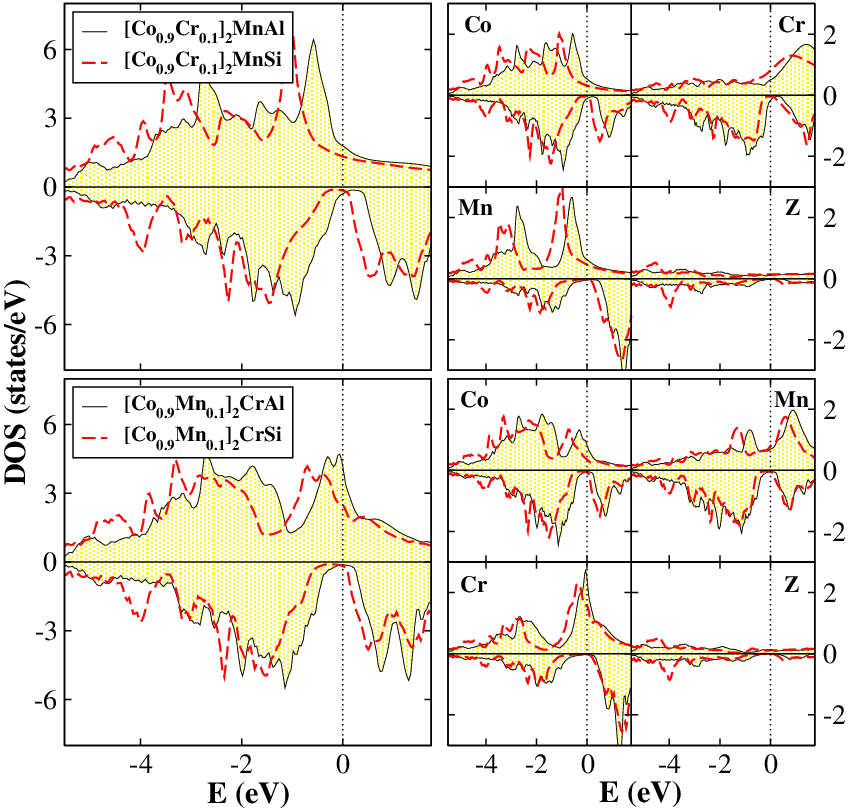}
\caption{(Color online) Total and atom-resolved DOS  for Cr
impurities in Co$_2$MnAl and Co$_2$MnSi alloys (upper panel) and
Mn impurities in  Co$_2$CrAl and Co$_2$CrSi compounds (lower
panel).} \label{fig1}
\end{minipage}
\hfil
\begin{minipage}[t]{.45\textwidth}
\includegraphics[width=\textwidth]{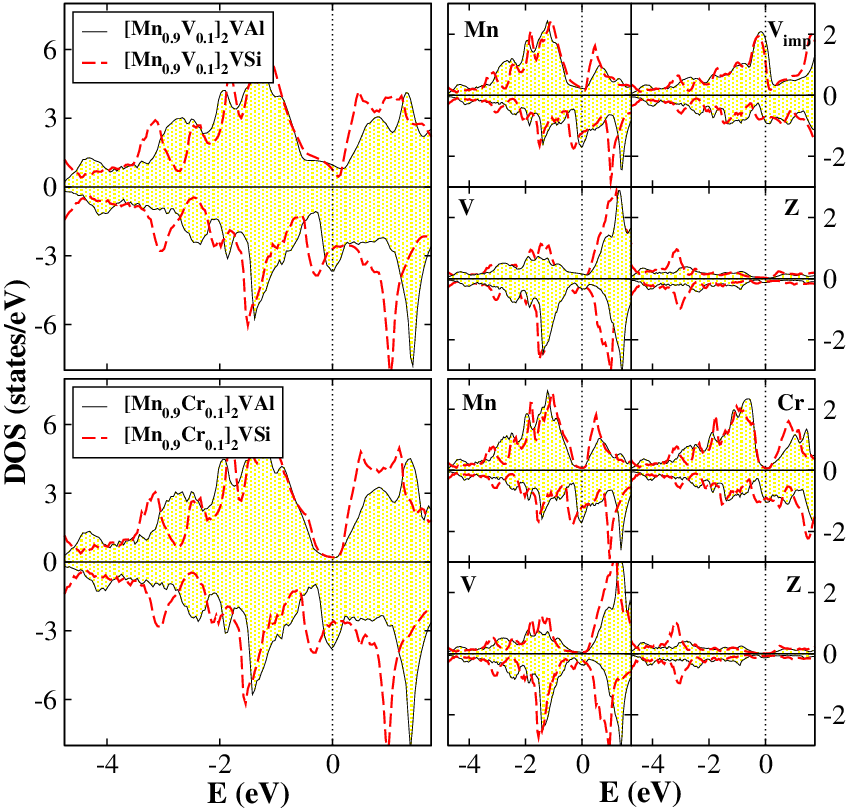}
\caption{(Color online) Total and atom-resolved DOS  for V (upper
panel) and Cr (lower panel) impurities in the case of Mn$_2$VAl
and Mn$_2$VSi alloys. DOS's have been scaled to one atom.  }
\label{fig2}
\end{minipage}
\end{figure}

In Ref. \cite{Mn2VZ3} we have shown that when we substitute Co or
Fe for Mn in the ferrimagnetic hafl-metallic Mn$_2$VAl and
Mn$_2$VSi alloys, the Co and Fe impurity atoms have spin moments
parallel to the V atoms at the Y site and antiparallel to the Mn
atoms. Thus the negative total spin moment of the alloys becomes
smaller in magnitude and for specific concentration of defects we
get a half-metallic antiferromagnet. We show now results when we
use V and Cr as impurity atoms which have lower valence than Mn.
As can be seen in Fig. \ref{fig2}, alloys with Cr impurities
almost keep unaltered the gap in the spin-up band for both Al and
Si-based compounds. Contrary to Cr atoms, the lighter vanadium
shows a smaller exchange splitting and the Fermi level falls
within a spin-up pick of the V DOS completely destroying the
half-metallicity. We should also note that we present results only
for a concentration of defects of $x$=0.1. We have performed
calculations also for $x$=0.025,0.05 and 0.2 but the behavior of
the compounds is similar to the case which we present.

Even more interesting is the behavior of the spin moments
presented in Table \ref{table}. In the case of Co and Fe
impurities, due to their positive spin moment, the total spin
moment was approaching zero as we increased the concentration.
Contrary to these chemical elements, V and Cr hybridize less with
the neighboring Mn atoms which show a more atomic like behavior
and thus increase the absolute value of their spin magnetic
moment. V or Cr atoms have smaller spin moments than Mn atoms but
they are in low concentration and as a result the total spin
moment increases in magnitude from the -2 $\mu_B$ of the perfect
Mn$_2$VAl and the -1 $\mu_B$ of the perfect Mn$_2$VSi (note that
the total spin moments are negative since these alloys have less
than 24 valence electrons according to the Slater-Pauling rule
\cite{Gala-Full}). Thus such a doping with Cr and V would have no
advantages over the perfect Mn$_2$VAl and Mn$_2$VSi compounds for
realistic applications.

\begin{figure}[htb]
\begin{minipage}[t]{.45\textwidth}
\includegraphics[width=\textwidth]{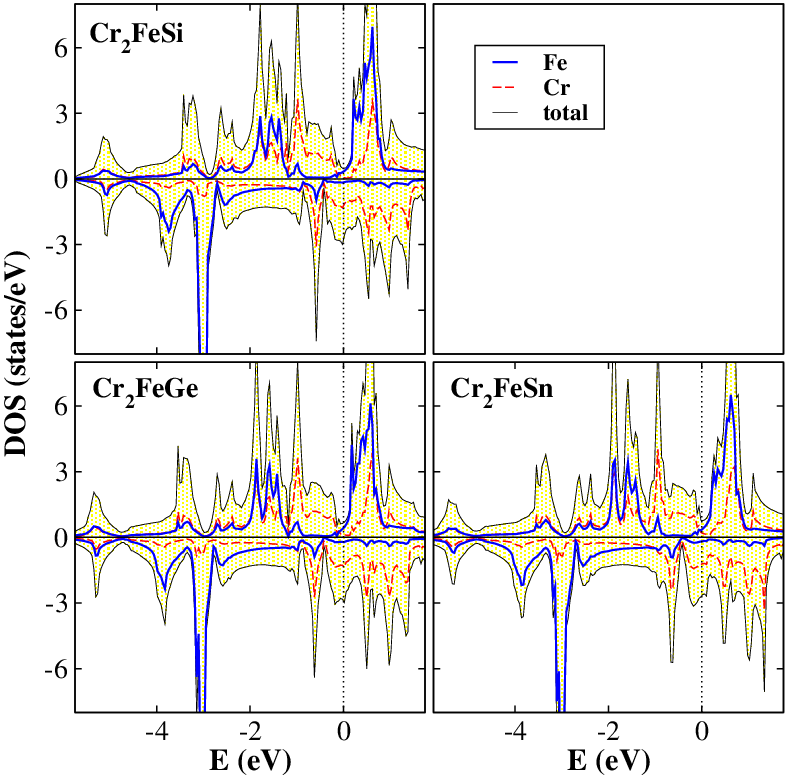}
\caption{Total, Cr- and Fe-resolved DOS for the Cr$_2$FeZ  alloys
where Z is Si, Ge or Sn and for a lattice constant of 6.2 \AA . }
\label{fig3}
\end{minipage}
\hfil
\begin{minipage}[t]{.45\textwidth}
\includegraphics[width=\textwidth]{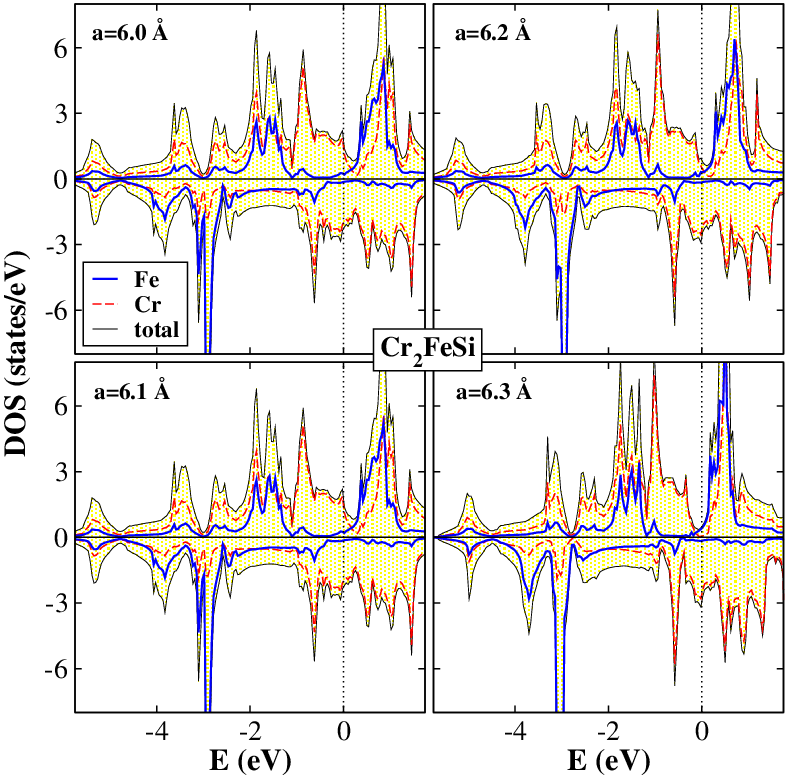}
\caption{(Color online) Total, Cr- and Fe-resolved DOS for the
Cr$_2$FeSi alloy and for four different values of the lattice
constant.} \label{fig4}
\end{minipage}
\end{figure}

\section{Cr$_2$FeZ (Z=Si,Ge,Sn) alloys :}\label{sec4}

Motivated by our results on the Cr$_2$MnZ alloys \cite{Cr2MnZ}, we
decided to study also another family of 24-valence electrons
compounds containing Fe instead of Mn: the Cr$_2$FeZ alloys where
Z is Si, Ge or Sn. In Fig. \ref{fig3} we present the total and the
Cr and Fe-resolved density of states (DOS) for a lattice constant
of 6.2 \AA\ for all three alloys. For this lattice constant all
the compounds under study present a region of low DOS in the
spin-up band and the Fermi level falls within this region.
Contrary to the compounds containing Mn, where the spin-up
occupied states are mainly of Cr character and the occupied
spin-down states are of mainly Mn character, for the Fe-based
alloys both the spin-up and spin-down occupied states exhibit a
more mixed character. The electronic structure is more complicated
than the Mn-based alloys and as a result when we slightly vary the
lattice constant the band-structure changes in a way that the the
region of low DOS is completely destroyed. This is illustrated in
Fig. \ref{fig4} where we present the DOS for Cr$_2$FeSi  for four
values of the lattice constant. Cr spin-up states below the Fermi
level and the Cr spin-up states just above the Fermi level move
one towards the other completely destroying the region of high
spin-polarization upon compression. Lattice constants larger than
6.3 \AA\ are unlikely to be achieved experimentally. The reason is
the different hybridization between the transition-metal atoms
when we substitute Fe for Mn. The smaller exchange splitting of
the Fe atoms closes the gap and this leads also to a different
distribution of the Cr charge which no more shows a large band-gap
as for the Mn-based alloys. Finally we should also discuss the
spin magnetic moments. Fe atoms  for the lattice constant of 6.2
\AA\ possess a moment slightly smaller than -3 $\mu_B$ and each Cr
atom has a spin moment of around 1.5 $\mu_B$ resulting in the
almost vanishing total spin moment requested by the Slater=Pauling
behavior for the half-metals with 24 valence electrons
\cite{Gala-Full}. Overall the Cr$_2$FeZ alloys with 24-valence
electrons are not suitable for realistic application, contrary to
the Cr$_2$MnZ compounds since half-metallicity is very fragile in
their case (e.g. calculations suggest that the region of low
spin-up DOS persists for Cr$_2$FeSi only between 6.2 and 6.3 \AA
).

\section{Summary :}\label{sec5}

We have presented first-principles calculations on several Heusler
alloys. Cr(Mn) impurites occupying Co sites in Co$_2$Mn(Cr)Al and
Co$_2$Mn(Cr)Si alloys couple antiferromagnetically to the other
transition metal atoms resulting in ferrimagnetic compounds with
lower total spin moments similarly to our previous studies on
these alloys \cite{PSS-RRL,unpublishedSSC}. V and Cr impurities
occupying Mn sites in ferrimagnetic Mn$_2$VAl and Mn$_2$VSi alloys
present spin moments parallel to the Mn atoms contrary to the
behavior of the Fe and Co impurities in these alloys and they lead
to larger absolute values of the total spin moments \cite{Mn2VZ3}.
Finally, rhe 24-valence electron alloys Cr$_2$FeZ (Z= Si, Ge, Sn)
show a region of low density of states instead of the real gap
presented by the isovalent Cr$_2$MnZ (Z= P, As, Sb, Bi)
\cite{Cr2MnZ} which persists only for a very narrow range of
lattice constants.

\end{document}